\documentclass[a4paper,12pt]{JHEP3}
\usepackage{amssymb}
\usepackage{mathrsfs}
\usepackage{cite}
\usepackage{texdraw}
\usepackage[english]{babel}
\usepackage{epsfig}

\csname @addtoreset\endcsname{equation}{section}

\newcommand{\bea}{\begin{eqnarray}}
\newcommand{\eea}{\end{eqnarray}}

\def\be{\beta}

\def\Om{\Omega}

\def\fft#1#2{{#1 \over #2}}

\title{Phase transitions and statistical mechanics for
 BPS Black Holes in AdS/CFT}
\author{Pedro J. Silva \\
Institut de Ciències de l'Espai (CSIC-IEEC) and
Institut de Fisica d'Altes Energies (IFAE),\\
E-08193 Bellaterra (Barcelona), Spain.\\
E-mail: \email{psilva@ifae.es}}

\abstract{Using the general framework developed in hep-th/0607056,
we study in detail the phase space of BPS Black Holes in AdS, for
the case where all three electric charges are equal. Although
these solitons are supersymmetric with zero Hawking temperature,
it turns out that these Black Holes have rich phase structure with
sharp phase transitions associated to a corresponding critical
generalized temperature. We are able to rewrite the gravity
variables in terms of dual CFT variables and compare the gravity
phase diagram with the free dual CFT phase diagram. In particular,
the elusive supergravity constraint characteristic of these Black
Holes is particulary simple and in fact appears naturally in the
dual CFT in the definition of the BPS Index. Armed with this
constraint, we find perfect match between BH and free CFT charges
up to expected constant factors.}


\begin{document}

\section{Introduction} \vspace{.5cm}

In \cite{Silva2} we developed a framework based on a multi-scaling
limit, that defines the "thermodynamics" o better "the statistical
mechanics" of supersymmetric solitons in gauge supergravity. One
of the basic ideas that grounds that work, is that a
supersymmetric partition function can be defined from the general
partition function as a combination of limits for the different
potentials, but not as the sole naive limit of "temperature
$\rightarrow 0$", since the BPS equation links all the different
charges. Once this framework was settled, as a result of the
combination of limits taken new conjugated potentials emerge
controlling the resulting BPS charges. These manipulations are
easy to implement in a supersymmetric field theory, and amazingly,
can also be implemented in supersymmetric configurations of gauge
supergravity. Then, as an application, using global and local
analysis we showed that BPS Black Holes (BH) present a phase
transition as a function of the generalized potentials. For
readers interested in the detail explanation of this framework, we
reefer to the original article.\vspace{.3cm}

\noindent In this letter, we continue our studies on statistical
mechanics properties of BPS supergravity solitons that were
started in a previous set of works \cite{Silva1,Silva2}. In
particular, we expand the above initial studies to describe the
phase diagram for BPS BH in AdS. Although we are in a
supersymmetric case, we find a sort of instability that translates
into a phase transition with a corresponding "generalized critical
temperature" \footnote{The above physics mimics the well known
Hawking-Page transitions at finite temperature of Schwarzschild
AdS BH and thermal AdS.}. Then, we connect our results with the
dual CFT picture, to search for a better understanding of the
microscopic structure of these BH. We found strong similarities
between the supergravity and the free CFT phase diagrams producing
a deeper understanding of the supergravity configuration and
constraints. \vspace{.3cm}

\noindent We make notice that the BPS BH studied here are just one
of the known families, that is chosen because is the only known
solution that has a well behaved off-supersymmetric extension. The
first solutions were found in
\cite{Gutowski:2004ez,Gutowski:2004yv} and the more general known
BPS solutions can be found in \cite{Kunduri:2006ek}. \vspace{.3cm}

\noindent Assuming that the AdS/CFT duality is correct
\cite{M1,Gubser:1998bc}, BH in AdS can be understood in the dual
CFT theory, as an ensemble of states at strong coupling. In fact,
the thermodynamical properties of the dual CFT have been
intensively studied for several years, in general the partition
function depends on the number of colors $N$, the canonical
ensemble used (like micro canonical, canonical or grand canonical)
and the coupling constant. The computation of the partition
function in the canonical ensemble at the free regime can be found
in \cite{Sundborg:1999ue}, extension to include the small couple
regime are presented in \cite{Aharony:2003sx}. Lately the
extensions to grand canonical ensemble to include R-charge
configurations can be found in \cite{Yamada:2006rx,Harmark:2006di}
and there are works in the literature where approximations of the
effective partition function at strong couplings are given
\cite{Basu:2005pj,Alvarez-Gaume:2006jg}. At last, in \cite{K}, the
supersymmetric partition function with all the relevant chemical
potentials turned on, was presented at zero coupling.\vspace{.3cm}

\noindent BPS BH by analogy, can be related to supersymmetric
ensembles at zero temperature but non-zero chemical potentials in
the dual CFT. These potential control the expectation value of the
pertinent conserved charges carried by the BH, like angular
momenta and electric charge. Unfortunately, it is not know how to
study the statistical mechanics properties of these ensembles in
the dual CFT theory at strong coupling, making very difficult the
comparison with the supergravity description. On the other hand,
it is possible to study the statistical mechanics of the free CFT
theory on a three-sphere at large $N$, where finite temperature
and BPS partition functions have been calculated (see for example
\cite{Sundborg:1999ue,Aharony:2003sx,K}). Therefore, {\it in this
note, we work out the strong coupling case using the BPS BH
soliton and then, we compare it with statistical mechanics studies
at zero coupling in the free CFT theory}.\vspace{.3cm}

\noindent From these studies, it is reported that there is an
amazing similarity between both dual frameworks. To be more
concrete, we obtained the BH phase diagram, showing the
corresponding phase transition and its interphase region. Also,
since is possible to define a generalized potential $w_+$
conjugated to the energy, we found useful to define a generalized
critical potential, as the minimal value of $w_+$ in the
interphase region.\vspace{.3cm}

\noindent One of the mayor puzzles for these BPS BH is that they
come with extra constraints (like extra relation between the
conserved charges above the BPS equality) that does not appears in
the dual CFT partition function. Somehow, BPS BH corresponds to a
particular kind of ensemble i.e. a hypersurface in the general
moduli space. Here, by writing the BH generalized potentials in
the "natural CFT basis", we discovered that the extra constraint
is very simple to write and also has a role in the CFT picture, in
the computation of the Index that counts supersymmetric states,
defined in \cite{K}.\vspace{.3cm}

\noindent Then, using the free CFT partition function together
with the newly found BH constraint, we obtained almost the same
phase diagram, and critical potential. The difference lies more in
the actual values than in the functional form. In fact, the
functional dependence of the resulting BH charges and CFT charges,
is the same up to constants in the case where we are well inside
the BH/deconfinement phase.\vspace{.3cm}

\noindent We would like to make notice that other studies at
finite temperature like \cite{Sundborg:1999ue} indicate that the
statistical mechanics on three-sphere has a smooth dependence on
the coupling. We believed that also in this case, on the top of
the BPS character of the sector under study this smooth dependence
reappears and is the underlying reason for the reported
similarities.\vspace{.3cm}

\noindent This work is organized as follows: In section \ref{gr}
we borrow from \cite{Silva2}, the necessary information to define
the relevant statistical mechanics studies of BPS BH. We then
elaborate further to define the generalized critical potential and
describe the phase diagram for this BPS sector. In section
\ref{weak}, we study the CFT dual ensemble by means of the free
BPS partition function together with the constraint found in the
previous section. Then, the corresponding critical potential and
phase space are described. Also the form of the charges in both
dual theories is compared well inside the BH/deconfinement phase.
In section \ref{end}, we comment the results, making some
conclusions and final remarks on future research and open
problems.

\section{The strong coupling case (supergravity)} \vspace{.5cm}

\label{gr}

We start this section with a short overview of the BH solutions of
minimal gauge supergravity in five dimensions of
\cite{Chong:2005hr}. In general, the solutions is characterized by
its energy $E$, two independent angular momenta $(J_1,J_2)$ and a
single electric charge $Q$. In the BPS regime, the solutions
preserve only a fraction of $1/16$ out of the total $32$
supercharges of the uplifted ten dimensional type IIB supergravity
and depending on the different range of values of its parameter
space, the solutions describe BPS BH or topological solutions with
no horizon (here we will concentrate in the BH case
only).\vspace{.3cm}

\noindent The form of the solution can be found in
\cite{Chong:2005hr} while in \cite{Silva2}, it is defined and
explicitly calculated the multi-scaling limit necessary to study
the statistical mechanics of the solution, in particular it is
showed how to define the BPS charges
$(Q_{bps},J^1_{bps},J^2_{bps})$, its generalized potentials
$(\phi,w_1,w_2)$ and the entropy $S_{bps}$. At the BPS bound the
different charges satisfy that
\[E_{bps}=\sqrt3\,Q_{bps}+J^1_{bps},+J^2_{bps}\,.\]
Here, to avoid a rather long discussion, we show the final
expressions referring to the original articles for details. First
we present the charges and entropy

\bea E_{bps}&=&\fft{\pi (a + b)(1-a)(1-b)+(1
+a)(1+b)(2-a-b)}{4(1 -a)^2 (1-b)^2}\,, \nonumber \\
J_{bps}^1&=&\fft{\pi (a +b) (2a +b + a b)}{4 (1-a )^2
(1-b)}\,,\quad J_{bps}^2=\fft{\pi (a +b) (a + 2b +
ab)}{4(1-a)(1-b)^2}\,,\nonumber \\
&&Q_{bps}=\fft{\sqrt{3}\pi(a+b)}{4(1-a)(1-b)}\,,\quad
S_{bps}=\fft{\pi^2(a +b)r_0}{2(1-a)(1- b)}\,. \nonumber \eea
\vspace{.3cm}

\noindent Second, the conjugated generalized potentials
\bea
w_1=\fft{\pi(1-a)(a+2ab+b^2+2b)}{r_{bps}(3r_{bps}^2+1+a^2+b^2)}\,,\quad
w_2=\fft{\pi(1-b)(b+2ab+a^2+2a)}{r_{bps}(3r_{bps}^2+1+a^2+b^2)}\,,\nonumber \\
\phi=\fft{\pi\sqrt{3}(a+b)(1-ab)}{r_{bps}(3r_{bps}^2+1+a^2+b^2)}\,.
\qquad\qquad\qquad\qquad\nonumber \eea
where $r_{bps}^2=a+b+ab$, and $(a,b)\leq1$. Notice that all the
above quantities come as a function of only two parameters
$(a,b)$.\vspace{.3cm}

\noindent These generalized potentials where defined in
\cite{Silva2}, as the next-to-leading term in the multi-scaling
limit that defines the BPS solution, of the well known potentials
of BH thermodynamics. In fact, the explicit definition is given by
\[ \be \rightarrow \infty\,, \quad \Om \rightarrow
\Om_{bps}-{w\over\be }+ O(\be^{-2})\,,\quad \Phi \rightarrow
\Phi_{bps}-{\phi\over\be }+ O(\be^{-2})\,. \]
where $(\beta,\Omega,\Phi)$ are respectively the inverse Hawking
temperature, angular velocity of the horizon and electric
potential of the general off-BPS BH solution under
study.\vspace{.3cm}

\noindent With these definitions we are ready to define the
Quantum Statistical Relation (QSR)
\bea I_{bps}=\phi\,
Q_{bps}+w_1J_{bps}^1+w_2J_{bps}^2-S_{bps}\nonumber \eea
where $I_{bps}$, is the value of the supersymmetric Euclidean
action of the corresponding BH. Solving for the explicit form of
these different quantities in the BH solution, we obtain
\bea I_{bps}=\fft{\pi^2(a+b)^2[-1+2b+b^2+b^2+a(2+5b+b^2)+
a^2(1+b)]}{4(1-a)(1-b)\sqrt{a+b+ab}(1+a^2+b^2+3(a+b+ab))}\,.
\nonumber \eea
\EPSFIGURE[ht]{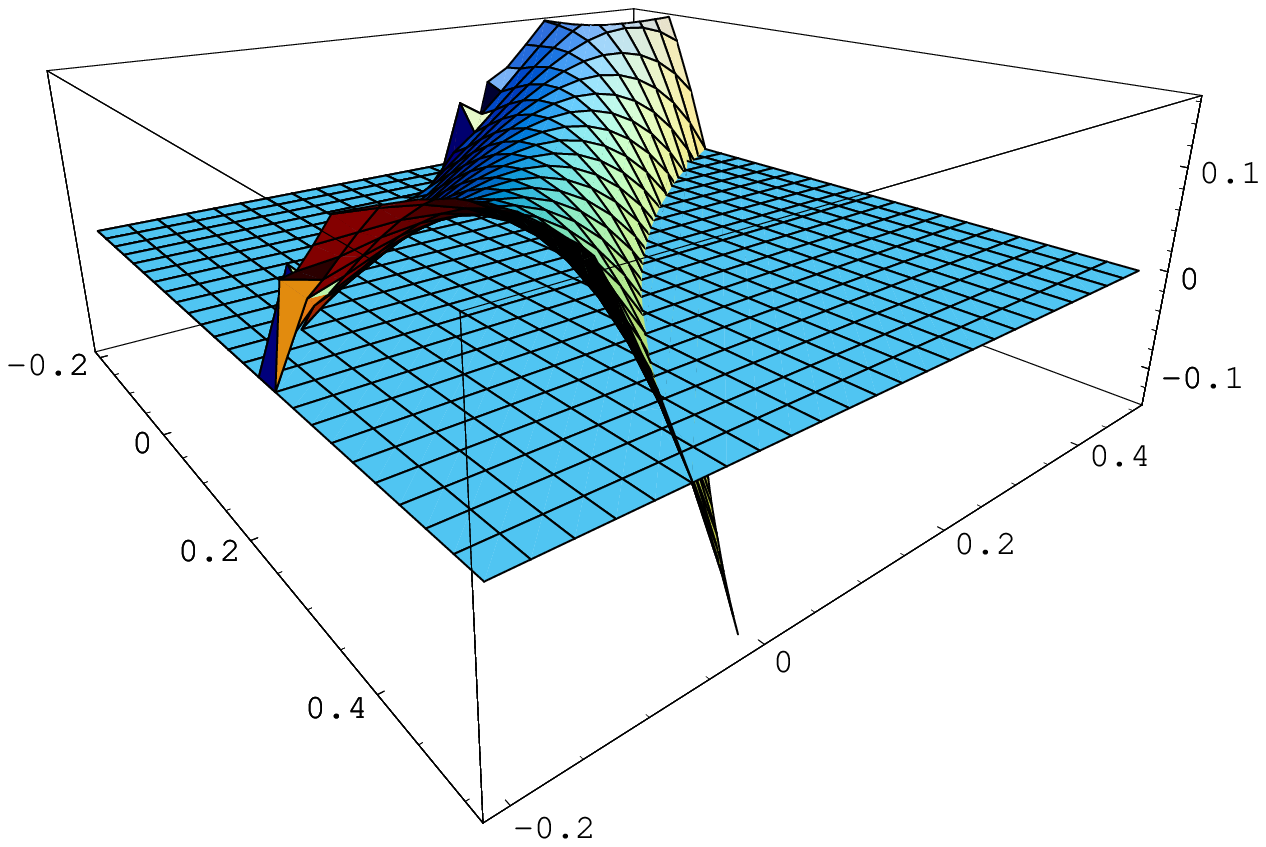}{Plot of the Euclidean action of the BPS BH
as a function of the parameters $(a,b)$. The flat plane
corresponds zero level surface.}
The range of the parameters $(a,b)$ is obtained by imposing that
the event horizon radius $r_{bps}$ and the energy are real
positive expression. In figure 1 and figure 2 we show two
different plots of $I_{bps}$, where in the first case, $I_{bps}$
is a function of $(a,b)$ while in the second $I_{bps}$ is at fixed
$b=.1$ and running $a$.
\EPSFIGURE[t]{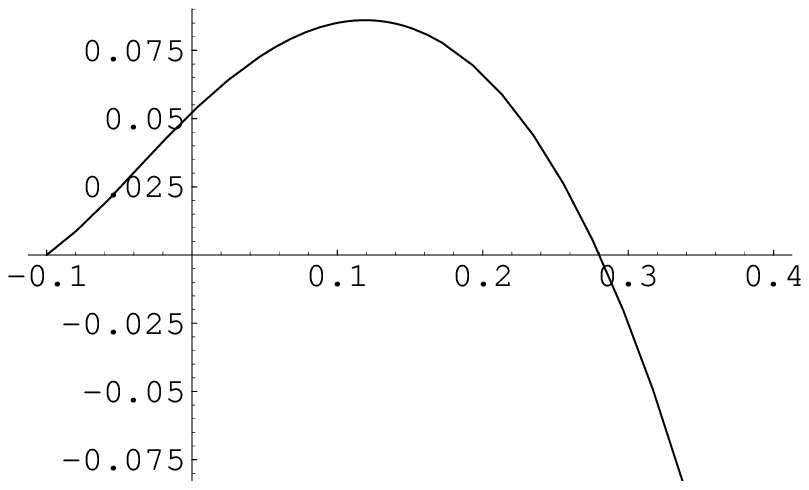,,scale=1.1}{Plot of the Euclidean action at
fixed $b=1/10$.}
\noindent As it was found in \cite{Silva2}, it is easy to see that
$I_{bps}$ is positive for small $(a,b)$ and negative for larger
values. Therefore we deduce that {\it Indeed there is a phase
transition}, where the BH solution is not any more the preferred
vacuum, but a meta-stable vacuum. The stable vacuum most probably
is a gas of superparticles in AdS, studied in detail in \cite{K}.
\vspace{.3cm}

\noindent At this point of the analysis, we found more convenient
to make a change of variables that facilitates the future
confrontation with CFT results of the next section. Then, we use
the following
\bea J^\pm=J_{bps}^1\pm J_{bps}^2 \quad,\quad
E=J_{bps}^1+J_{bps}^2+\sqrt3 Q_{bps} \quad,\quad Q=Q_{bps}\nonumber \\
w_\pm={(w_1\pm w_2)\over 2}\quad,\quad \lambda=(\phi-\sqrt3
w_+)\quad,\quad S=S_{bps}\,. \quad\quad \nonumber \eea
In this new set of variables the QSR is given by
\bea I_{bps}=w_+\, E + w_- J^- + \lambda\,Q-S\,.\nonumber \eea
Some of the advantages of these new variables is that we have
obtained a potential conjugated to the energy (a sort of
generalized inverse of the temperature) $w_+$ and that the
Euclidean action in not a functional of $J^+$. Also, we get the
new left an right angular momenta $(J^+,J^-)$ and the generalized
potentials $(w_-,\lambda)$ conjugated to $(J^-,Q)$ respectively.
\vspace{.3cm}

\noindent Notice that, in principle we should have three
independent parameters (four charges $(E,J^+,J^-,Q)$ modulo the
BPS bound). Instead our expressions come as functions of only two
parameters, showing that these BH are constraint systems. We have
found that the corresponding constraint is amazingly simple in
this set variables, namely that
\[ \lambda(a,b)=-{w_+(a,b)\over \sqrt3}\,.\]
In other words, we are looking into BH solutions where two out of
the three generalized potentials are proportional! In essence, we
have only two degrees of freedom, that we will choose to be
$(w_+,w_-)$ for the rest of the analysis with no lose of
generality.\vspace{.3cm}

\noindent Next, to better characterized the above phase
transition, we first identify its locus, where $I_{bps}=0$. After
some algebra the above requirement reduces to a quadratic equation
with the two associated roots
\[ a(b)_\mp=\frac{(-b^2+5b+2)\mp\sqrt{b^4+2b^3+13b^2+16b+8}}{2(1+b)}\,. \]

\noindent We found that $a(b)_-$ is not a physical solution since
in this range of parameters $(a(b)_-,b)$, the event horizon radius
$r_{bps}$ is pure imaginary. The other solution $a(b)_+$ is
physical and in fact corresponds to the phase transition observed
before.
\EPSFIGURE[t]{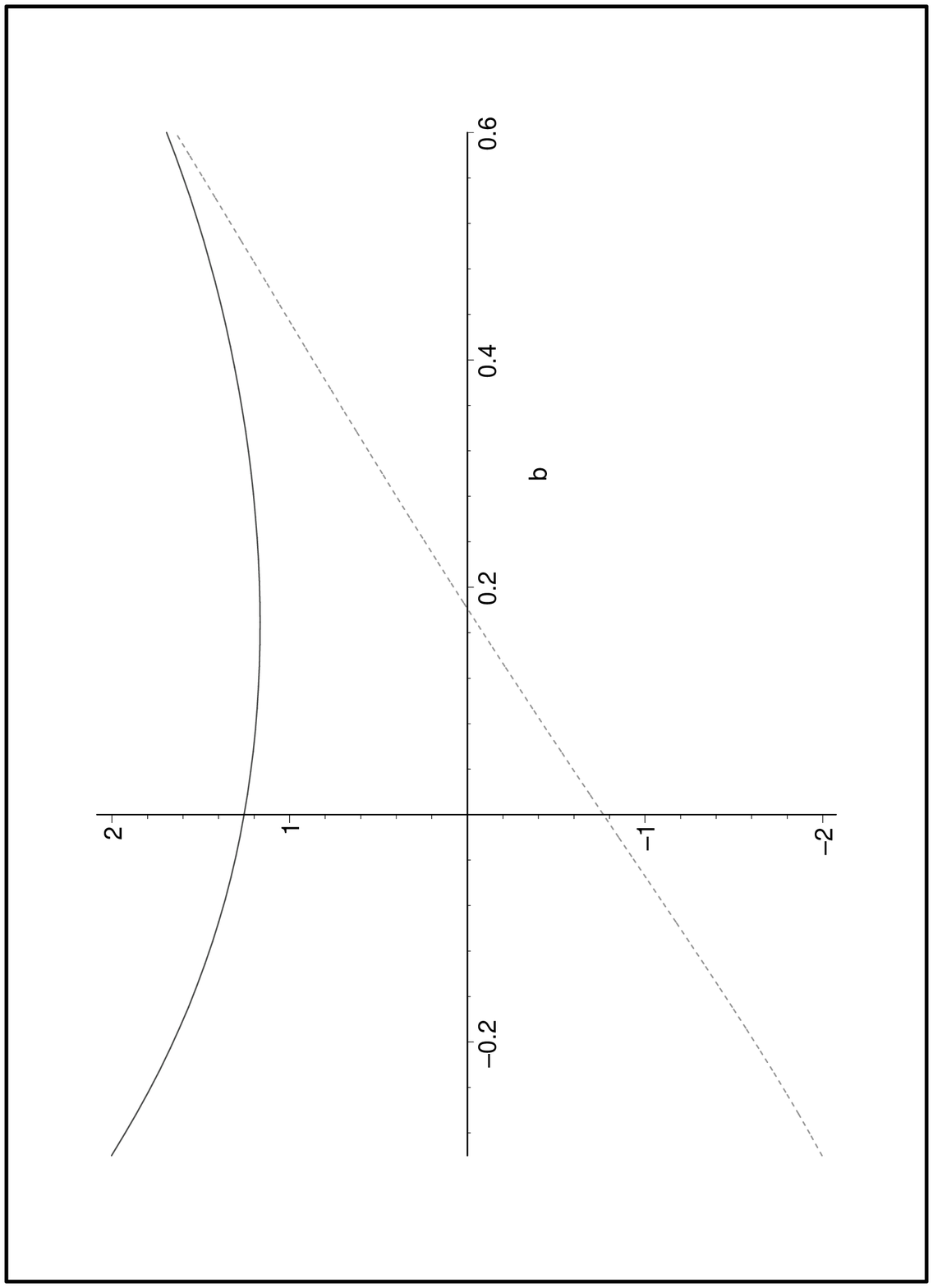,angle=-90,scale=.5}{Plot of $w_+$ and $w_-$
at the phase transition, as a function of $b$. $w_+$ is plotted
with a continuous line, while $w_-$ is in dotted line.}
Then, in figure 3 we show a plot of the two generalized potentials
at the phase transition locus as a function of the parameter $b$.
It is not difficult to evaluate numerically all three generalized
potentials $(w_+,w_-,\lambda)$, at the maximum of the critical
generalized temperature (minimum of $w_+$), obtaining
\bea w_+=1.1668 \quad,\quad \lambda=-0.673 \quad,\quad w_-=0
\,.\label{strong}\eea
These values correspond to the point where $a=b\approx 0.1813$.
That $w_+$ is at its minimum when $w_-$ is zero, is expected since
we are looking for the maximal generalized critical temperature.
In fact, at this point there is no $J^-$ charge and we need more
energy i.e. less $w_+$ to obtain a phase transition. \vspace{.3cm}

\noindent Unfortunately, we could not solve for $b$, to rewrite
the phase diagram in terms of $(w_+,w_-)$ alone in a analytic
form. Nevertheless, after some inspection it is clear that $w_-$
is almost a linear function of $b$ in the vicinity of the minima
for $w_+$. Based on this observation we used the approximation
that $w_-\approx 4.3b-.766$ to solve for $b$ and then draw the
final version of the phase diagram figure 4.
\EPSFIGURE[t]{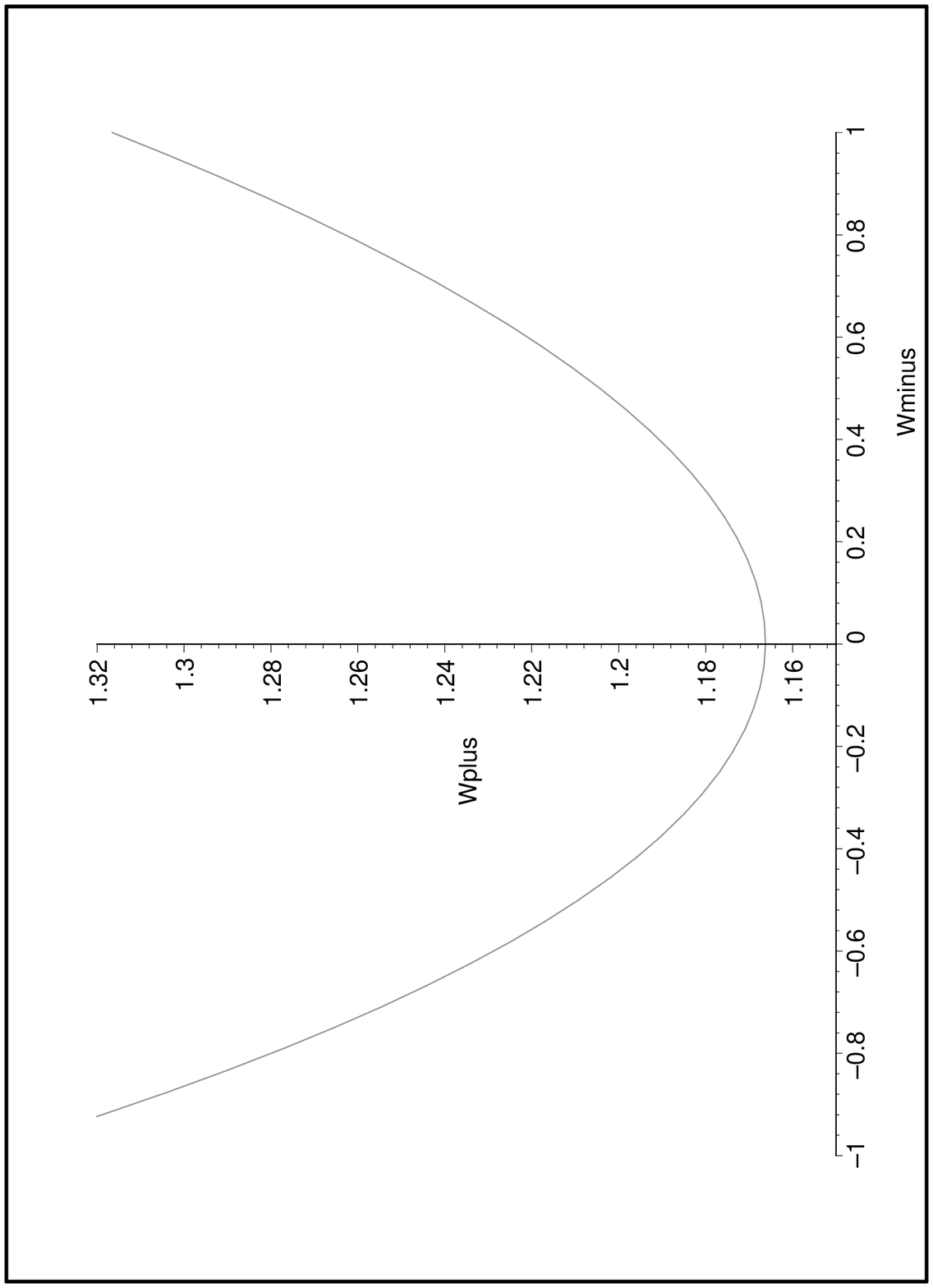,angle=-90,scale=.5}{Plot of $w_+$ as a
function of $w_-$, the curve shows the phase transition.}
In this plot, it can clearly be seen the phase transition diagram
as a function of the two generalized potentials $(w_+,w_-)$
conjugated to the two charges $(E,J^-)$. The region in the
exterior of the curve, corresponds the BH phase, while the region
in the interior of the curve corresponds to the AdS phase.

\section{The zero coupling case (free CFT)} \vspace{.5cm}

\label{weak}

In this section, we study the same supersymmetric sector of the
above section, but this time in the dual picture using the
conformal field theory language. The ideal situation would be to
calculate the CFT partition function in the strong coupling regime
and then compare its structure against the supergravity results.
Unfortunately, such calculations are not within our actual
capabilities and therefore we have decided to work instead with
the simpler case where the partition function is known,
corresponding to the free theory i.e. at zero coupling.
\vspace{.3cm}

\noindent Before starting the actual calculations, let us recall
that similar studies at finite temperature resulted in a
celebrated connection between the Hawing-Page phase transition
(Schwarzschild AdS BH and thermal AdS) and the
deconfinement/confinement transition of large $N$, strongly
coupled $\mathcal{N}$=4 Super Yang Mills theory on a three sphere
\cite{Witten:1998qj,Witten:1998zw}. In the present case, we will
work in the grand canonical ensemble with all possible potentials
turned on (since we are interested in configurations with
generically non-zero expectation value for angular momenta and
electric charge), focussing on the partition function over BPS
states that is associated to zero temperature statistical
mechanics. \vspace{.3cm}

\noindent As it is well known, $\mathcal{N}$=4 Super Yang Mills at
zero coupling on a three-sphere is simple enough for explicit
calculations. We are dealing with free dynamics in the presence of
a global constraint of neutral color in all states (related to
Gauss law on compact manifolds). In \cite{K}, the form of the
partition function relevant to our studies was computed (see also
\cite{Sundborg:1999ue,Aharony:2003sx} for general calculations at
non-zero temperature). If we start with the definition of the
supersymmetric partition function, constrained to the case where
all three R-charges are set equals \footnote{We are considering
only the case of all three R-charges equals i.e. $q=q_1=q_2=q_3$.
Also we have changed by little the original notation of \cite{K},
in order to accommodate better our previous section.}, we first
write
\begin{equation} Z=\sum_{bps} e^{-\beta e + 3\mu q + 2\xi j_-}\,,
\label{Z}\end{equation}
that defines the conjugated potentials $(\beta,\mu,\xi)$ to the
charges $(e,q,j_-)$ respectively energy, R-charge and angular
momentum. Then it can be shown that
\bea Z=\int {DU \exp {\left\{
\sum_n\left[f_B(x^n,y^n,v^n)+(-1)^{n+1}f_F(x^n,y^n,v^n)\right]{Tr
U^nU^{-n}\over n}\right\}}}\,,\nonumber \label{partition}\eea
where $U$ is a unitary matrix, $x=e^{\beta}\,, y=e^\mu \,,
v=e^{\xi}\,,$
\bea f_B={3yx+x^2\over (1-xv)(1-x/v)}\,,\quad
f_F={x^{3/2}\left[(v+1/v)y^{3/2}+3y^{1/2}\right]-x^{5/2}y^{3/2}\over
(1-xv)(1-x/v)}\,,\nonumber \eea
In \cite{K}, it was found that the above partition function
undergoes a phase transition at finite values of the generalized
potentials, where one phase is independent of $N$, while the other
phase goes like $N^2$.\vspace{.3cm}

\noindent The phase transition can be studied searching for the
singular behavior of the partition function $Z$. The locus of the
phase transition (i.e. the generalized critical surface), is found
by the strongest singularity of $Z$ i.e.
\bea f_B(x^n,y^n,v^n)-(-1)^{n+1}f_F(x^n,y^n,v^n)=1\,,
\label{z}\eea
that corresponds to the case $n=1$.\vspace{.3cm}

\noindent At first sight, it is easy to see that our partition
function depends on too many variables to successfully reproduce
the supergravity results. Recall that the supergravity BH depends
on only two parameters and hence would corresponds to a particular
class of ensemble within the general ensemble of (\ref{Z}).
Therefore, we have to find somehow a constraint to reduce the
number of independent generalized potentials from three to
two.\vspace{.3cm}

\noindent At present, the nature of this constraint is by no means
clear in the CFT picture (see \cite{simon} for some clues).
Nevertheless in the previous section we learned how the
supergravity generalized potentials where related by the equation
$\lambda=-w_+/ \sqrt 3$. Therefore, we found natural to impose
this relation upon the CFT potential to check its implication and
results. To implement this constraint in the CFT picture, we just
have to notice that the normalization of the potential $\mu$ and
its charge $q$ is of, by a factor of $\sqrt 3$ when compared to
the supergravity definitions. Once this is taken into account the
constraint in CFT variables reads
\begin{equation}\mu=-\beta/3 \label{mu} \end{equation}
Note that this is the same relation found in \cite{K}, for the
definition of the Index. Off course, in the partition function
$Z$, there is no cancellation between configurations due to the
extra factor $(-1)^F$ characteristic of the Index.\vspace{.3cm}

\noindent Therefore with this constraint the locus of interphase
is defined by
\[ {\left[3x^{4\over 3}+(1+v+1/v)x^{2}
+3x^{10\over6}-x^{3}\right]\over (1-xv)(1-x/v)}=1\,.
\]
At this point, it is not difficult to find the value of the
minimal generalized critical potential $\beta$ and the
corresponding values of the other two potentials,
\bea \beta\approx1.6301 \quad,\quad \mu\approx-0.5435 \quad,\quad
\xi=0 \,.\label{free}\eea
Using the constraint partition function, we can in principle
obtain the phase diagram as a function of $(\beta,\xi)$. For
technical reasons we found more easy to solve for $\xi$ as a
function of $\beta$, and then obtain numerically the desiderate
free large $N$ CFT phase diagram. The resulting phase diagram is
showed in figure 5.
\EPSFIGURE[ht]{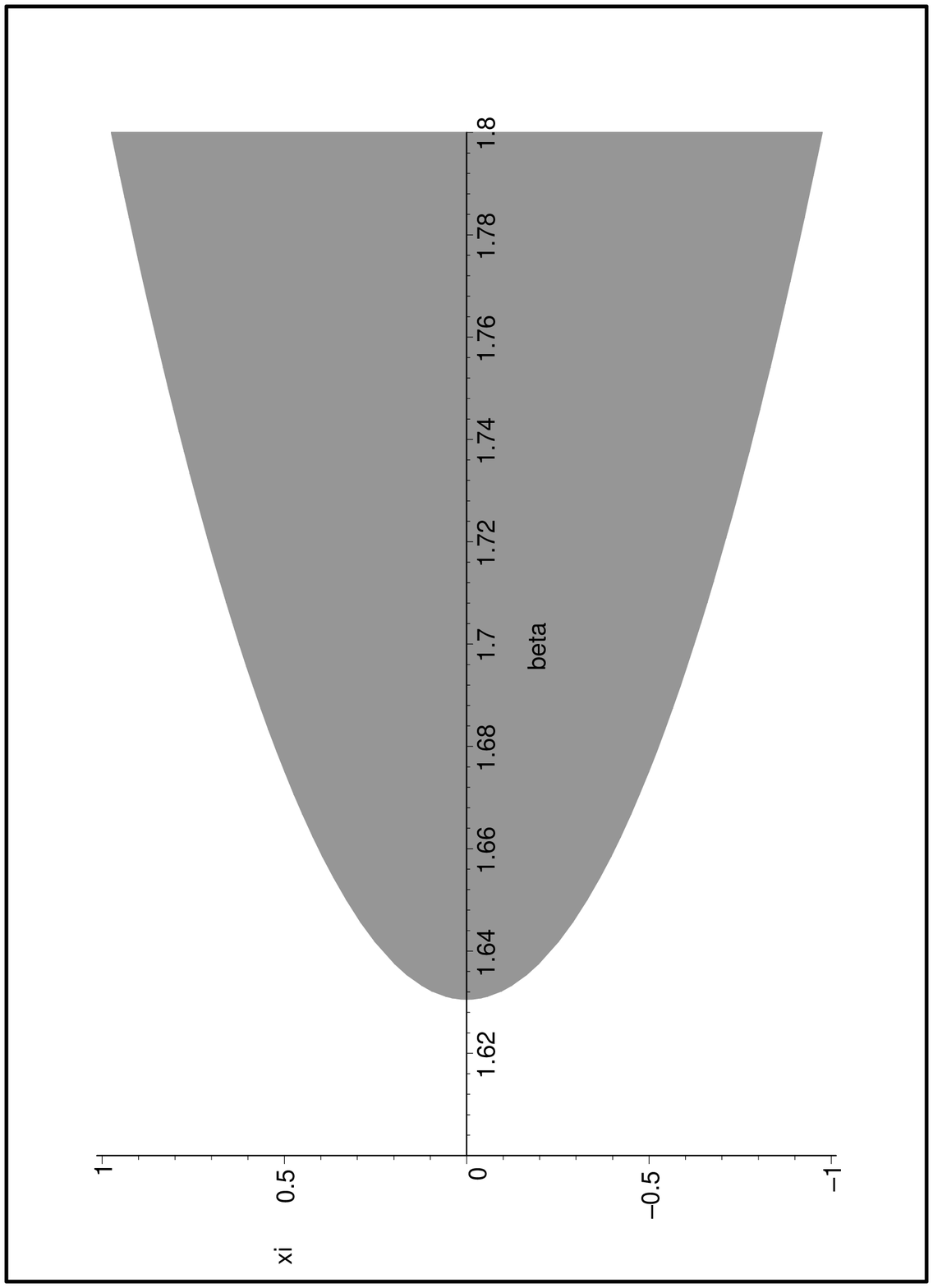,angle=-90,scale=.5}{Plot of the phase space
with the line of interface. In the plot, $\xi$ runs along the
vertical axes, while $\beta$ runs in the horizontal axes. The
shaded region corresponds to the region where $Z$ is independent
of $N$.}\vspace{.3cm}

\noindent Note that the shape of the diagram matches very well the
supergravity case, showing that the free theory calculation is a
valuable regime where to look for BH physics. Regarding the
generalized critical potential, note that the free value is bigger
than the strongly coupled value and this can be explain since in
the free regime, there are "more" BPS states than in the strongly
couple regime and therefore we need less "temperature" to create a
BH out of a thermal ensemble.\vspace{.3cm}

\noindent It is interesting to cross check our results with an
observation made in \cite{K}, where the expectation values of
$(e,j_-,q)$ were used to guess a constraint in the particular
regime of the moduli space well inside the BH phase. There, it was
found $\mu\approx -0.504$. Here, we can use this value of $\mu$
into eqn. (\ref{z}) to calculate the corresponding generalized
Hagedorn potential $\beta$ (assuming therefore a constant $\mu$ as
a rough approximation). The result is $\beta\approx 1.65$ at
$\xi=0$ and hence gives a value that is still larger than the
strong coupling result of eqn. (\ref{strong}) but also larger than
our zero coupling result of eqn. (\ref{free}). Therefore
(\ref{free}) is closer to the strong coupling value that is
consistent with the idea that our constrain is more
accurate.\vspace{.3cm}

\noindent Another important check is to compare the form of the
resulting charges in both dual descriptions. In general points of
the phase space, the CFT expressions for the charges are too
complicated but in \cite{K}, it was found that things get more
manageable if we are well inside the BH phase. There, it was used
that $(\beta,\xi)\ll 1$, while $\mu$ was left free.\vspace{.3cm}

\noindent In this regime, the explicit expressions for the energy,
electric charge, angular momentum and entropy (respectively
($e,q,j_-,s$)) are
\bea &&e={2\beta f(\mu)N^2\over (\beta^2-\xi^2)^2}\quad ,\quad
2j_-={2\xi f(\mu) N^2\over
(\beta^2-\xi^2)^2}\,,\nonumber \\
&&q={g(\mu)N^2\over (\beta^2-\xi^2)}\quad ,\quad s={(3f(\mu)-\mu
g(\mu)) N^2\over (\beta^2-\xi^2)}\,,\eea
where $f(\mu)=(\zeta(3)+3Pl(3,y)-3Pl(3,-y^{1/2})-Pl(3,-y^{3/2}))$,
$g(\mu)=\partial_\mu f/3$, $Pl(s,z)$ is the PoliLog function and
$\zeta(n)$ is the Riemann's Zeta function. Notice now, that if we
use our constraint (\ref{mu}), $\mu\ll 1$, and therefore the above
expression become even more simple reducing to
\bea &&e={14\zeta(3)\beta N^2\over (\beta^2-\xi^2)^2}\quad ,\quad
2j_-={14\zeta(3)\xi N^2\over
(\beta^2-\xi^2)^2}\,,\nonumber \\
&&q={\pi^2 N^2\over 4(\beta^2-\xi^2)}\quad ,\quad s={21\zeta(3)
N^2\over (\beta^2-\xi^2)}\,,\eea
where we have used that $f(0)=7\zeta(3)$ and
$g(0)=\pi^2/2$.\vspace{.3cm}

\noindent This results are to be compared with the BH charges in a
corresponding region of the parameter space, namely where
$a=1-(\tilde{\beta}+\tilde{\xi})$ and
$b=1-(\tilde{\beta}-\tilde{\xi})$. The final expressions for the
charges is
\bea &&E={8\tilde\beta N^2\over
(\tilde\beta^2-\tilde\xi^2)^2}\quad ,\quad 2J_-={8\tilde\xi
N^2\over
(\tilde\beta^2-\tilde\xi^2)^2}\,,\nonumber \\
&&Q={\sqrt{3}N^2\over (\tilde\beta^2-\tilde\xi^2)}\quad ,\quad
S={2\pi\sqrt3 N^2\over (\tilde\beta^2-\tilde\xi^2)}\,,\eea
where the factor $\sqrt3$ of $Q$ is due to the different
normalization between $Q$ and $q$. Notice that both relations are
functionally identical and the only difference lies in the value
of the different constants. The fact that the CFT and the dual BH
expression are so similar for the case $\mu\ll 1$ was noted in
\cite{K}, the difference is that here we obtain this limit as the
necessary condition once the constraint \ref{mu} is used.
\vspace{.3cm}

\noindent Therefore, in spite all the difference that characterize
the supergravity and the free CFT frameworks, we have found in
this work strong similarities between the two dual descriptions
even so they are calculated at such different regimes on the
coupling constant. We suspect that the match of the structure is
not only linked to supersymmetry but may also indicate that the
statistical mechanics of the CFT on the three-sphere, has a smooth
dependence on the coupling constant.

\section{Discussion}
\label{end}

In this work, we have studied the statistical mechanics properties
of BPS BH of minimal gauge supergravity in five dimensions. In
order to carry on these studies, we used the new framework defined
in \cite{Silva2}. Then, based on the AdS/CFT duality, we contrast
our results with CFT calculations using more standard statistical
mechanics methods developed in \cite{K}. \vspace{.3cm}

\noindent As main conclusion we point out that these BH present a
rich phase structure with phase transitions and generalized
critical potential. The results tell us about the phase structure
of the dual CFT theory at strong coupling. Also, we found that the
free theory has strong similarities with the strong coupling case,
since in this regime, the CFT presents the same king of phase
diagrams and only constant factors seems to be different.
\vspace{.3cm}

\noindent This work does not exhaust all the physical structure of
the BPS phase space. In particular we have only compared the
supergravity results with the free CFT picture, but it is well
known the things may change in the interacting theory even a very
weak coupling. Therefore we think it will be vary interesting to
study the weakly and strongly coupled supersymmetric CFT partition
function (see \cite{Aharony:2003sx,Yamada:2006rx,Harmark:2006di,
Alvarez-Gaume:2006jg,Basu:2005pj,Harmark:2006ta} for some
extension in the non-supersymmetric case along this
directions).\vspace{.3cm}

\noindent At a more technical level, we found that the BH ensemble
is characterized by a simple constraint, acting on the generalized
potentials, that also appears in the CFT theory. Unexpectedly, in
the CFT theory this constraint shows up only in the construction
of an Index and not in the partition function \cite{K}.
Apparently, these BH obey strict relations of group theory
representation that we believe should be utilized to guess
properties of the corresponding microstates. Nevertheless, at
present we have not a clear picture of the above. \vspace{.3cm}

\noindent It will be very interesting to study more general BPS
Black holes or/and possibly Black Rings, but unfortunately all the
known solutions are either singular in the BPS regime or not known
out of the BPS regime. As a consequence of the above, both cases
we can not be studied with the framework of \cite{Silva2}, since
we need well behaved solutions to consider the multi-scaling
limit. We are currently working in a generalization to cover these
singular cases too \cite{Silva3}. \vspace{.3cm}

\noindent Also, we point out that in \cite{K} it was found another
type of phase transition more alike to a Bose-Einstein
condensation in more supersymmetric sectors of the CFT. It will be
very interesting to see what is the dual description of such
phenomena in supergravity. \vspace{.3cm}

\noindent At last, in retrospective, this work may be seen as a
confirmation that the general framework developed in \cite{Silva2}
is correct and therefore should be useful to better understand the
microscopic structure of BH and quantum gravity.

\section*{Acknowledgments}

\small The author would like to thanks Daniela Zanon, and in
general the string theory group at Milano for the excellent
atmosphere found in the last years, that helped and facilitated my
research.

\noindent This work was partially funded by the Ministerio de
Educacion y Ciencia under grant FPA2005-02211 and by INFN, MURST
and by the European Commission RTN program HPRN-CT-2000-00131.
\normalsize


\end{document}